\documentclass[12pt,a4paper,oneside]{article}

\usepackage[T1]{fontenc}
\usepackage[utf8]{inputenc}

\usepackage{hyperref}
\usepackage{fancyhdr}
\usepackage{xargs} 
 \usepackage[nomargin]{fixme}

\usepackage{setspace}

\usepackage{authblk} 

\usepackage{newlfont}
\usepackage{amssymb}
\usepackage{amsmath}
\usepackage{latexsym}
\usepackage{amsthm}
\usepackage[mathscr]{eucal}
\usepackage{mathrsfs}
\usepackage{fourier}
\usepackage{siunitx}

\usepackage{graphicx}

\usepackage{caption}
\usepackage{subcaption}

\DeclareUnicodeCharacter{2002}{}
\theoremstyle{plain}                    %stile corsivo
\newtheorem{teo}{Theorem}[section]      %definizione ambiente teorema
\newtheorem*{teo*}{Theorem} %definizione ambiente teorema non numerato
\newtheorem*{prop*}{Proposition}    %definizione ambiente proposizione non numerata
\theoremstyle{definition}               %stile roman
\newtheorem{claim}{Claim}[section]

\theoremstyle{remark}                   %stile per osservazioni
\title{Late time evolution of negatively curved FLRW models}
\author[1,2]{Roberto Giamb\`{o}\thanks{roberto.giambo@unicam.it}}

\author[3]{John Miritzis\thanks{imyr@aegean.gr}}

\author[1,4]{Annagiulia Pezzola\thanks{annagiulia.pezzola@studenti.unicam.it, a.pezzola1@unimc.it}}

\affil[1]{School of Science and Technology, Mathematics Division, University of Camerino (Italy)}
\affil[2]{INFN, Sezione di Perugia, 06123-Perugia (Italy)}
\affil[3]{Department of Marine Sciences, University of the Aegean (Greece)}
\affil[4]{PhD School, University of Macerata (Italy)}

\begin{document}
\maketitle
                %va in una pagina nuova
%%%%%%%%%%%%%%%%%%%%%%%%%%%%%%%%%%%%%%%%%non numera l'ultima pagina sinistra
%\clearpage{\pagestyle{empty}\cleardoublepage}

%\pagenumbering{roman}                   %serve per mettere i numeri romani

%\tableofcontents
%\newpage\null\thispagestyle{empty}\newpage
                 %serve per mettere i numeri romani
                                  
%\listoftodos
% \newpage 

\begin{abstract}
We study the late time evolution of negatively curved
Friedmann--Le\-ma\^{\i}tre--Robert\-son--Walker (FLRW) models with a perfect fluid
matter source and a scalar field nonminimally coupled to matter. Since, under mild
assumptions on the potential $V$, it is already known 
-- see e.g. \cite{gimi} -- that equilibria corresponding to
non-negative local minima for $V$ are asymptotically stable, we classify all
cases where one of the energy components eventually dominates. In particular
for nondegenerate minima
with zero critical value, we rigorously prove that if $\gamma$, the parameter
of the equation of state is larger than $2/3$, then there is a transfer of
energy from the fluid and the scalar field to the energy density of the scalar
curvature. Thus, the scalar curvature, if present, has a dominant effect on
the late evolution of the universe and eventually dominates over both the
perfect fluid and the scalar field. The analysis in complemented with the case where $V$ is exponential and therefore the scalar field diverges to infinity. 
\end{abstract}                

\section{Introduction}

Scalar fields implement a useful tool used by theorists for the description of
the early inflationary phase and of the present accelerating expansion of the
Universe, \cite{K3,wein2,K11}. Scalar fields arise in several conformally
equivalent theories of gravity, e.g., in higher order gravity theories (HOG),
in string theories \cite{gasp} and in scalar-tensor theories involving scalar
field self-interactions and dynamical couplings to matter \cite{fuma}. An
important example of a scalar field coupled to matter is provided by the
general form of scalar-tensor theories of gravity \cite{fuma,K14,lsf}, where
the action in the Einstein frame takes the form
\begin{equation}
S=\int d^{4}x\sqrt{-g}\left\{  R-\left[  \left(  \partial\phi\right)
^{2}+2V\left(  \phi\right)  \right]  +2\chi^{-2}L_{\mathrm{m}}\left(
\widetilde{g}_{\mu\nu},\Psi\right)  \right\}  , \label{scte}%
\end{equation}
where%
\[
\widetilde{g}_{\mu\nu}=\chi^{-1}g_{\mu\nu},
\]
and $\chi=\chi\left(  \phi\right)  $ is the coupling function; matter fields
are collectively denoted by $\Psi$. This action contains as special cases HOG
theories with $\chi\left(  \phi\right)  =e^{\sqrt{2/3}\phi}$ and
$\widetilde{g}_{\mu\nu}=e^{-\sqrt{2/3}\phi}g_{\mu\nu}$. Non minimally coupling
occurs also in models of chameleon gravity \cite{khwe,wate}, with
$\widetilde{g}_{\mu\nu}=e^{2\beta\phi}g_{\mu\nu},$ where $\beta$ is a coupling
constant. The same form of coupling has been proposed in models of the so
called coupled quintessence \cite{amen1} (see also \cite{psc,lumu,bico} for
more general couplings).

Variation of the action (\ref{scte}) with respect to the metric $g$ yields the
field equations,
\begin{equation}
G_{\mu\nu}=T_{\mu\nu}\left(  g,\phi\right)  +T_{\mu\nu}^{\mathrm{m}}\left(
g,\Psi\right)  , \label{confm}%
\end{equation}
where $T_{\mu\nu}^{\mathrm{m}}$ is the matter energy momentum tensor. The
Bianchi identities imply that the total energy-momentum tensor is conserved
and therefore there is an energy exchange between the scalar field and
ordinary matter. In all the above examples, the conservation of their sum is
provided by the equations (compare to \cite{amen1}),
\begin{equation}
\nabla^{\mu}T_{\mu\nu}^{\mathrm{m}}\left(  g,\Psi\right)  =QT^{\mathrm{m}%
}\nabla_{\nu}\phi,\ \ \ \ \nabla^{\mu}T_{\mu\nu}\left(  g,\phi\right)
=-QT^{\mathrm{m}}\nabla_{\nu}\phi, \label{bian}%
\end{equation}
where $Q:=d\ln\chi/d\phi,$ depends in general on $\phi$ and $T^{\mathrm{m}}$
is the trace of the matter energy-momentum tensor, i.e., $T^{\mathrm{m}%
}=g^{\mu\nu}T_{\mu\nu}^{\mathrm{m}}\left(  g,\Psi\right)  $. Variation of $S$
with respect to $\phi$ yields the equation of motion of the scalar field,%
\begin{equation}
\square\phi-\frac{dV}{d\phi}=-QT^{\mathrm{m}}. \label{emsf}%
\end{equation}

In early investigations in scalar-field cosmology a minimal coupling of the
scalar field was assumed, (see for example the review articles \cite{K11,lsf}
and references therein). It is true that inclusion of non minimal coupling
increases the mathematical difficulty of the analysis; however, it is
important to consider non minimal coupling in scalar field cosmology
\cite{fuma}. Many physical theories predict the presence of a scalar field
coupled to matter and therefore, the introduction of non minimal coupling is
not a matter of taste \cite{fara1}. Models with exponential potentials have
been intensively studied not only because of the variety of alternative
theories of gravity which predict exponential potentials, but also due to the
fact that this potential has the nice property that $V^{\prime}\propto V$
which allows for the introduction of normalized variables according to the
formalism of Wainwright et al \cite{wael} -- see however,
\cite{bkgz,K9,gimi,dkzpct}. Another large class of potentials used in
scalar-field cosmological models has a local minimum. In view of the unknown
nature of the scalar field supposed to cause accelerated expansion, it is
important to investigate the general properties shared by all
Friedmann--Lema\^{\i}tre--Robertson--Walker (FLRW) models with a scalar field
irrespective of the particular form of the potential.

In this paper we study the late time evolution of initially expanding
negatively curved FLRW models with a scalar field having an arbitrary bounded
from below potential function $V\left( \phi\right) $. Although most of the
literature deals with the flat case, negatively curved FLRW models cannot be
ruled out by current observation in principle, and have been already analysed
i.e. in \cite{nmc} for the massless case (see also final section
\ref{sec:final} below). Our results are rigorously proved and do not depend on
the specific form of the potential function, but possibly at most on its local
form near its minimum. Ordinary matter is described by a barotropic fluid with
equation of state
\begin{equation}
p=(\gamma-1)\rho,\ \ \ 0<\gamma< 2. \label{eq:eos}%
\end{equation}
The scalar field is nonminimally coupled to matter according to (\ref{scte}),
(\ref{confm}) and (\ref{emsf}). Under general assumptions on the potential
function $V(\phi)$ we study the late time mutual behavior of the energies
associated to the scalar and the fluid, and their relation with the
\textquotedblleft energy\textquotedblright\ associated to the spatial
curvature of the cosmological model. In particular, in case of a nondegenerate
minimum of the potential with vanishing critical value, we show that for
$\gamma<2/3$ the perfect fluid eventually dominates the energy density of the
scalar field, i.e., $\Omega_{\rho}\rightarrow1$, $\Omega_{\phi},\,\Omega
_{k}\rightarrow0$, a property already shared by flat models for $\gamma<1$.
However, if $\gamma>2/3$, the energy density of the scalar curvature
eventually dominates over both the perfect fluid and the scalar field, i.e.,
$\Omega_{k}\rightarrow1\ $and $\Omega_{\phi},\,\Omega_{\rho}\rightarrow0$
asymptotically. This result shows that the scalar curvature, if present, has a
dominant effect on the late evolution of the universe.

The paper is organized as follows. Section \ref{sec:model} briefly recollects
the basic ideas of the model studied. The results on the asymptotic behavior
of the energy are described in Section \ref{sec:conti}. In paragraph
\ref{sec:nondeg} we consider the case when the scalar approaches a local
nondegenerate minimum of the potential $V(\phi)$, and find in Theorem
\ref{final} the spectrum of the qualitative behavior of the energies in terms
of the parameter $\gamma$ entering the equation of state of the fluid. To
complete the analysis in analogy with \cite{gimi}, we also present in
paragraph \ref{sec:exp} the case of the exponential potential, finding out
that the curvature energy may take over the other energies also in this case.
The final Section \ref{sec:final} is devoted to conclusions and perspectives.

\section{Model description}

\label{sec:model} The metric is given by the general, possibly nonflat FLRW
metric
\[
\text{d}s^{2}=-\text{d}t^{2}+a(t)^{2}\bigg(\frac{\text{d}r^{2}}{1-kr^{2}%
}+r^{2}\text{d}\Omega^{2}\bigg),
\]
where $\text{d}\Omega^{2}=\text{d}\vartheta^{2}+\sin^{2}{\vartheta}%
\text{d}\varphi^{2}$. For this metric, assuming that $L_{m}$ is the Lagrangian
of a perfect fluid with equation of state \eqref{eq:eos}, the field equations
(\ref{confm}) reduce to the Friedmann equation,
\begin{equation}
H^{2}+\frac{k}{a^{2}}=\frac{1}{3}\left(  \rho+\frac{1}{2}\dot{\phi}%
^{2}+V\left(  \phi\right)  \right)  , \label{Friedmann}%
\end{equation}
and the Raychaudhuri equation,
\begin{equation}
\dot{H}=-\frac{1}{2}\dot{\phi}^{2}-\frac{\gamma}{2}\rho+\frac{k}{a^{2}},
\label{RR}%
\end{equation}
while the equation of motion of the scalar field (\ref{emsf}), becomes
\begin{equation}
\ddot{\phi}+V^{\prime}(\phi)+3H\phi+Q(\phi)\frac{(3\gamma-4)}{2}\rho=0,
\label{Mo}%
\end{equation}
where $Q(\phi)$ is the logarithmic derivative of the coupling function
$\chi(\phi)$ which is supposed to be strictly positive and differentiable. The
Bianchi identities (\ref{bian}) yield the conservation equation,
\begin{equation}
\dot{\rho}+3\gamma\rho H=Q(\phi)\frac{4-3\gamma}{2}\rho\dot{\phi}.
\label{conssfjm}%
\end{equation}
Here, $a\left(  t\right)  $ is the scale factor, an overdot denotes
differentiation with respect to time $t,$ $H=\dot{a}/a$ and units have been
chosen so that $c=1=8\pi G$. The potential $V\left(  \phi\right)  $ of the
scalar field is a $C^{2}$ function and $dV/d\phi$ is denoted by $V^{\prime
}\left(  \phi\right)  $.

Setting $y=\dot{\phi}$ and $\alpha(\phi):=\tfrac{4-3\gamma}{2}Q(\phi)$, we
obtain the system
\begin{align}
\dot{\phi}  &  =y,\nonumber\\
\dot{y}  &  =-3Hy-V^{\prime}\left(  \phi\right)  +\alpha\rho,\nonumber\\
\dot{\rho}  &  =-3\gamma\rho H-\alpha\rho y,\label{Sistema}\\
\dot{H}  &  =-\frac{1}{2}y^{2}-\frac{\gamma}{2}\rho+\frac{k}{a^{2}},\nonumber
\end{align}
subject to the constraint
\begin{equation}
3H^{2}+\frac{3k}{a^{2}}=\rho+\frac{1}{2}y^{2}+V\left(  \phi\right)  .
\label{cons1}%
\end{equation}

We begin with some general properties of the system (\ref{Sistema}) with the
constraint (\ref{cons1}). Firstly, the system shares the remarkable property
of the Einstein equations that, if equation (\ref{cons1}) is satisfied at some
initial time, then it is satisfied throughout the evolution. Secondly,
following the arguments in \cite{fost, miri03}, one can show that an initially
expanding flat or negatively curved universe remains ever-expanding. Thirdly,
the third of (\ref{Sistema}) implies that the set $\rho=0$ is invariant.
Therefore, if initially $\rho$ is positive, it remains positive for ever.
Furthermore, for $k=0,-1$ the fourth equation of (\ref{Sistema}) implies
$\dot{H}<0$, thus $H$ is a decreasing function of time $t$ and is bounded from
below either by $0$ or $\sqrt{V(\phi_{\ast})/3})$ where $V^{\prime}(\phi
_{\ast})=0$. We observe that the function $W(t)$ defined as
\begin{equation}
W(t)=W(\phi(t),y(t),\rho(t),H(t))=H^{2}-\frac{1}{3}\left(  \frac{1}{2}%
y^{2}+V(\phi)+\rho\right)  , \label{eee}%
\end{equation}
satisfies the equation
\begin{equation}
\dot{W}=-2HW, \label{W}%
\end{equation}
therefore, $\mathrm{sgn}(W)$ is invariant under the flow of \eqref{Sistema}and
since
\[
W(t)=-\frac{k}{a^{2}(t)}%
\]
it must be $k=-\mathrm{sgn}(W(0))$.

Using the constraint (\ref{cons1}) to eliminate $a$, we observe that the
critical points of \eqref{Sistema} are given by $(\phi=\phi_{\ast}%
,y=0,\rho=0,H=\pm\sqrt{V(\phi_{\ast})/3})$ where $V^{\prime}(\phi_{\ast})=0$.

Following \cite{gimi}, we assume two properties for the scalar potential
$V(\phi)$: (i) the (possibly empty) set $\left\{  \phi:V(\phi)<0\right\}  $ is
bounded, and (ii) the set of critical points is finite.
Under these assumptions one can show -- see e.g. \cite[Proposition 1]{gimi} -- that $(\phi_{\ast
},y_{\ast}=0,\rho_{\ast}=0,H_{\ast}=\sqrt{V(\phi_{\ast})/3})$, where
$\phi_{\ast}$ is a -- possibly degenerate -- strict local minimum for the
potential $V(\phi)$, is an asymptotically stable equilibrium point for
expanding cosmologies in the open spatial topologies $k=0$ and $k=-1$. The
purpose of this paper is to establish some results on the asymptotic behaviour
in the case $k=-1$.

%\begin{prop*}
%\cite[Proposition 1]{gimi}
%\label{thm:stab}
%Let $\phi_{\ast}$ a strict local minimum for $V(\phi)$, possibly degenerate,
%with nonnegative critical value. Then, $\mathbf{p}_{\ast}=(\phi_{\ast}%
%,y_{\ast}=0,\rho_{\ast}=0,H_{\ast}=\sqrt{V(\phi_{\ast})/3})$ is an
%asymptotically stable equilibrium point for expanding cosmologies in the open
%spatial topologies $k=0$ and $k=-1$.
%\end{prop*}

\section{Asymptotic behavior of the energy}

\label{sec:conti} In the following, we are going to study the late time
behaviour of solutions of (\ref{Sistema}), which are initially expanding,
i.e., $H(0)>0$. Our aim is to study which is the asymptotically dominating
energy in the above model: we have the energy $\rho$ associated to the perfect
fluid and the energy associated to the scalar field
\begin{equation}
\epsilon=\frac{1}{2}y^{2}+V(\phi), \label{eq:eps}%
\end{equation}
so that using equation \eqref{eee} we have
\begin{equation}
\label{identita}\Omega_{\rho}+\Omega_{\phi}+\Omega_{k}=1
\end{equation}
where
\begin{equation}
\Omega_{\rho}=\frac{\rho}{3H^{2}},\quad\Omega_{\phi}=\frac{\epsilon}{3H^{2}%
},\quad\Omega_{k}=\frac{W}{H^{2}}=\frac{1}{a^{2}H^{2}}, \label{eq:normen}%
\end{equation}
are the normalized energies related to the perfect fluid, the scalar field and
the spatial scalar curvature. Notice that the third component may become
relevant, unlike the flat case $k=0$ where the leading contribution is either
given by the perfect fluid or the scalar. We will consider in subsection
\ref{sec:nondeg} the case of a nondegenerate minimum of the potential and show
the main result of this paper, Theorem \ref{final}, whereas in subsection
\ref{sec:exp} we will take into account the exponential potential with
critical point at infinity.

\subsection{Nondegenerate minimum of the potential}

\label{sec:nondeg}

Let us study the asymptotic behaviour when the scalar field approaches a local
minimum of the potential $V(\phi)$. We will focus on the case when
$V(\phi_{\ast})=0$ -- indeed, when $V(\phi_{\ast})>0$, the energy density of
the scalar field approaches a strictly positive value and eventually
dominates. The less trivial case is given when the critical value for the
potential is zero, and we consider the case where this minimum is
\textit{nondegenerate}. Furthermore, we assume without loss of generality that
$\phi_{\ast}=0$ and therefore, the potential near its minimum takes the form
\begin{equation}
\label{eq:quadr-pot}V(\phi)=\frac{1}{2}\lambda^{2}\phi^{2}+O(\phi
^{3}),\ \ \ \lambda>0.
\end{equation}
From now on, the higher order terms in $V(\phi)$ will be systematically
neglected, since it can be shown that the results we are going to state are
not affected.

We recall \cite[Theorem 2]{gimi} that, in the flat case $k=0$, $\Omega_{\rho}$
eventually dominates when $\gamma<1$, whereas $\Omega_{\phi}$ eventually
dominates when $\gamma>1$ in a generic way, i.e. except at most for a
particular solution of the system. Let us go and see what happens when $k=-1$
and, in principle, another form of\ \textquotedblleft energy\textquotedblright%
\ deriving from the spatial curvature of the cosmological model enters into play.

%\begin{teo*}
%\cite[Theorem 2]{gimi}
%\label{thm:main-old}
%Let $\phi_{\ast}$ be a nondegenerate minimum of $V(\phi)$ with zero critical
%value. Consider the solutions of \eqref{Sistema} with $k=0$ approaching the
%(asymptotically stable) equilibrium point $(\phi_{\ast},y=0,\rho=0,H=0)$. Then
%if $\gamma<1$, for every such solution the fluid energy $\rho$ eventually
%dominates over the scalar energy $\epsilon$, whereas if $\gamma>1$, $\epsilon$
%eventually dominates over $\rho$ in a generic way, i.e., except at most for a
%particular solution of the system.
%\end{teo*}

%In the following, we will extend this study to the nonflat case $k=-1$.
Using {\eqref{Friedmann}} we can eliminate $k$ from \eqref{Sistema} thereby
obtaining%
\begin{align}
\dot{\phi}  &  =y,\nonumber\\
\dot{y}  &  =-3Hy-\lambda^{2}\phi+\alpha\rho,\label{S1}\\
\dot{\rho}  &  =-3\gamma\rho H-\alpha\rho y,\nonumber\\
\dot{H}  &  =-\frac{1}{3}y^{2}+\left(  \frac{1}{3}-\frac{\gamma}{2}\right)
\rho+\frac{1}{6}\lambda^{2}\phi^{2}-H^{2}.\nonumber
\end{align}
The phase space of the system is the set
\begin{equation}
\left\{  \left(  \phi,y,\rho,H\right)  \in\mathbb{R}^{4}:H^{2}>\frac{1}%
{3}\left(  \rho+\frac{1}{2}y^{2}+V(\phi)\right)  \right\}  , \label{dis}%
\end{equation}
due to the constraint (\ref{cons1}).

To tackle the problem we consider expansion-normalized variables, a
traditionally useful approach for flat cosmologies (see e.g. \cite{K9}) that
will be exploited here also for the $k=-1$ case under examination. We set
\begin{equation}
w=\frac{\lambda}{\sqrt{6}}\frac{\phi}{H},\ \ \ \ z=\frac{1}{\sqrt{6}}\frac
{y}{H},\ \ \ \ u=\frac{\sqrt{W}}{H}, \label{eq:normalized}%
\end{equation}
in order to write normalized energies \eqref{eq:normen} as
\begin{equation}
\Omega_{\rho}=1-(u^{2}+w^{2}+z^{2}),\quad\Omega_{\phi}=w^{2}+z^{2},\quad
\Omega_{k}=u^{2}, \label{eq:energyuwz}%
\end{equation}
and the system \eqref{S1} takes the form
\begin{align}
\dot{w}  &  =\lambda z+wH\left(  3z^{2}+u^{2}+\frac{3}{2}\gamma\left(
1-w^{2}-z^{2}-u^{2}\right)  \right)  ,\nonumber\\
\dot{z}  &  =-\lambda w+zH\left(  -3+3z^{2}+u^{2}\right)  +3H\left(
1-w^{2}-z^{2}-u^{2}\right)  \left(  \frac{\gamma}{2}z+\frac{\alpha}{\sqrt{6}%
}\right)  ,\label{eqn:b}\\
\dot{u}  &  =-uH\left(  1-3z^{2}-u^{2}-\frac{3}{2}\gamma\left(  1-w^{2}%
-z^{2}-u^{2}\right)  \right)  ,\nonumber\\
\dot{H}  &  =-H^{2}\left(  3z^{2}+u^{2}+\frac{3}{2}\gamma\left(  1-w^{2}%
-z^{2}-u^{2}\right)  \right)  .\nonumber
\end{align}
The above system is slightly complicated, with respect to its flat counterpart, by the introduction of one more equation, the third one, that in the flat $k=0$ is trivially satisfied since $W=0$ and then $u=0$. The main complication arising will be that, in the argument below, we will have to employ spherical coordinates instead of polar ones. 

We also observe that, 
in view of the first of \eqref{eq:normalized}, and recalling that $\alpha$ is in principle a function of the scalar field $\phi$ only, we can consider $\alpha
=\alpha(w,H)$; 
however one should bear in mind that we are considering solutions such that
$\phi\rightarrow\phi^{\ast}$, and then $\alpha$ will approach a constant as
$t\rightarrow+\infty$. This fact will be used in the proofs throughout below.

Finally we notice that, unlike the case that will be treated in next paragraph
\ref{sec:exp}, the presence of the terms $\lambda z$ and $-\lambda w$ in the
first two equations of \eqref{eqn:b} does not allow to consider, as is usually
done for similar studies, a normalized time $\tau$, and therefore the first
three equations no more decouple from the fourth one, as happens in the
exponential case. 

As mentioned above, it will be useful to consider the system in spherical
coordinates $(R,\theta,\eta)$ defined as%
\begin{align}
u  &  =R\cos{\eta,}\nonumber\\
w  &  =R\sin{\eta}\cos{\theta,}\label{eq:polaruwz}\\
z  &  =R\sin{\eta}\sin{\theta,}\nonumber
\end{align}
and the system is rewritten in the following form:%
\begin{subequations}
\begin{align}
\dot{\theta}+\lambda &  =3H\cos{\theta}\left(  -\sin{\theta}+\frac{\alpha
}{\sqrt{6}\sin{\eta}}\frac{\left(  1-R^{2}\right)  }{R}\right)
,\label{eq:stud1}\\
\dot{\eta}  &  =H\cos{\eta}\left(  \sin{\eta}\left(  1-3\sin^{2}{\theta
}\right)  +\frac{3\alpha\sin{\theta}}{\sqrt{6}}\frac{\left(  1-R^{2}\right)
}{R}\right)  ,\label{eq:stud2}\\
\dot{R}  &  =H\left(  1-R^{2}\right)  \left(  R\left(  \sin^{2}{\eta}\left(
1-3\sin^{2}{\theta}\right)  -1+\frac{3\gamma}{2}\right)  +\frac{3\alpha
\sin{\eta}\sin{\theta}}{\sqrt{6}}\right)  ,\label{eq:stud3}\\
\dot{H}  &  =H^{2}\left(  R^{2}\left(  \sin^{2}{\eta}\left(  1-3\sin
^{2}{\theta}\right)  -1\right)  -\frac{3\gamma}{2}\left(  1-R^{2}\right)
\right)  , \label{eq:stud4}%
\end{align}
compare with \cite[eq.(23)]{gimi}. Notice that, using \eqref{eq:energyuwz} and
\eqref{eq:polaruwz}, the normalized energies now are given by
\end{subequations}
\begin{equation}
\Omega_{\rho}=1-R^{2},\,\quad\Omega_{\phi}=R^{2}\sin^{2}{\eta},\quad\Omega
_{k}=R^{2}\cos^{2}{\eta}, \label{eq:norm-pol}%
\end{equation}

The main result of this paper is the following:

\begin{teo}
\label{final} Let $\phi_{\ast}$ be a nondegenerate minimum of $V(\phi)$ with
zero critical value. Consider the solutions of \eqref{Sistema} with $k=-1$
approaching the (asymptotically stable) equilibrium point $(\phi_{\ast
},y=0,\rho=0,H=0)$. Then if $\gamma<2/3$ then the normalized energy of the
perfect fluid dominates asymptotically:
\[
\Omega_{\rho}\to1,\qquad\Omega_{\phi},\,\Omega_{k}\to0.
\]
On the other side, when $\gamma>2/3$, the energy of the scalar curvature
dominates asymptotically:
\[
\Omega_{k}\to1,\qquad\Omega_{\phi},\,\Omega_{\rho}\to0.
\]

\end{teo}

Observe that, in particular, the energy associated to the scalar field never
eventually dominates\footnote{Also observe that the transition case
$\gamma=\tfrac{2}{3}$ is excluded from the current analysis, similarly to the
transition case $\gamma=1$ of \cite[Theorem 2]{gimi}}.

To show the above result we consider the compact and positively invariant set
$\mathcal{U}=\left\{  u^{2}+w^{2}+z^{2}\leq1,u\geq0\right\}  \times\left\{
H\in\lbrack0,H_{0}]\right\}  $. By LaSalle's theorem the possible $\omega
$--limit points for trajectories living in $\mathcal{U}$ are given by the
circles $w^{2}+z^{2}=r_{\infty}^{2}$ ($=$constant) with $u=u_{\infty}$
constant. We must then observe that a trajectory of the system can admit only
one such $\omega$--limit circle, otherwise with a simple contradiction
argument one would obtain infinite limit circles of this kind. Therefore, with
reference to the variable change given by \eqref{eq:polaruwz}, we find that
$\exists\lim_{t\to+\infty} R(t)=:R_{\infty}\ge0$ and, if $R_{\infty}>0$, also
$\exists\lim_{t\to+\infty}\eta(t)=\eta_{\infty}\in\mathbb{R}$. Then Theorem
\ref{final} is a straightforward consequence of the following two claims, that
will be shown to hold.

\begin{claim}\label{thm:cl1}
If $R_{\infty}=0$ then $\gamma<2/3.\smallskip$.
\end{claim}

\begin{claim}\label{thm:cl2}
If $R_{\infty}>0$ and $\gamma\neq2/3$ then $\gamma>2/3$ and
$R_{\infty}=1.$
\end{claim}

Claim \ref{thm:cl1} holds because, if $R\rightarrow0$, then from {\eqref{eq:stud4}} we get
that $\dot{H}\approx-\frac{3\gamma}{2}H^{2}$, and integrating twice we obtain
$a\approx t^{\frac{2}{3\gamma}}.$ Since $\dot{a}\approx\frac{2}{3\gamma
}t^{\frac{2}{3\gamma}-1},$ then it must be
\[
\Omega_{k}=R^{2}\cos^{2}{\eta}=u^{2}=\frac{W}{H^{2}}=\frac{1}{\dot{a}^{2}%
}\approx\frac{9}{4}\gamma^{2}t^{2\left(  1-\frac{2}{3\gamma}\right)  },
\]
which recalling $R\rightarrow0$ is consistent only when $\gamma<2/3$.

To prove Claim \ref{thm:cl2} we begin by showing by contradiction that $\sin\eta_{\infty
}=0$. Indeed if that is not the case, 
since $H$ monotonically goes
to zero, from {\eqref{eq:stud1} we obtain
$\theta\approx-\lambda t$ 
and consequently the following asymptotic estimates as $t\rightarrow+\infty
$ are found:
\[
\int_{0}^{t}H\cos{\eta}\sin{\eta}\left(  1-3\sin^{2}{\theta}\right)
\,\mathrm{d}s=\int_{0}^{t}H\cos{\eta}\sin{\eta}\left(  \frac{3}{2}\cos
{2\theta}-\frac{1}{2}\right)  \,\mathrm{d}s\approx\int_{0}^{t}-\frac{1}%
{2}H\cos{\eta_{\infty}}\sin{\eta_{\infty}}\,\mathrm{d}s+c_{0}%
\]
and
\[
\int_{0}^{t}H\cos\eta\frac{3\alpha\sin{\theta}}{\sqrt{6}}\frac{\left(
1-R^{2}\right)  }{R}\,\mathrm{d}s\approx c_{1}\in\mathbb{R},
\]
for some $c_{0},c_{1}\in\mathbb{R}$. Using the above estimates in
\eqref{eq:stud2} we get
\[
\eta(t)\approx-\frac{1}{2}\cos{\eta_{\infty}}\sin{\eta_{\infty}}\ln
a(t)+c_{2},
\]
for some $c_{2}\in\mathbb{R}$, and therefore it must be $\cos\eta_{\infty}=0$
and so $\sin^{2}\eta_{\infty}=1$. This implies the following asymptotic
estimate in {\eqref{eq:stud4}}:
\[
\int_{0}^{t}\frac{\dot{H}}{H^{2}}\text{d}s\approx-\nu t,
\]
where $\nu=\left(  \frac{3}{2}R_{\infty}^{2}+\frac{3}{2}\gamma(1-R_{\infty
}^{2})\right)  $. Integrating once again we find that
\[
u^{2}=\frac{1}{\dot{a}^{2}}\approx t^{2(1-1/\nu)}%
\]
and recalling that $u^{2}=R^{2}\cos^{2}\eta\rightarrow0$ it must be $\nu<1$,
that implies $\gamma<2/3$ and $R_{\infty}<\sqrt{2/3}$. But similar arguments
applied to \eqref{eq:stud3} give
\[
R(t)-R(0)\approx\int_{0}^{t}H(1-R_{\infty}^{2})\frac{3}{2}R_{\infty}%
(\gamma-1)\mathrm{d}s+c_{3}%
\]
for some $c_{3}\in\mathbb{R}$, and then either $\gamma=1$ or $R_{\infty}=1$,
that are both inconsistent with what we have found above. Then $\sin
\eta_{\infty}=0$. Now, recalling that we are excluding $\gamma=2/3$,
\eqref{eq:stud3} again produces the asymptotic estimate
\[
\frac{\dot{R}}{R\left(  1-R^{2}\right)  }\approx H\left(  \frac{3}{2}%
\gamma-1\right)  ,
\]
that implies $R/\sqrt{1-R^{2}}\approx a^{\frac{3}{2}\gamma-1},$ which is
consistent with $R_{\infty}>0$ only if $\gamma>2/3$ and $R_{\infty}=1$,
finally proving Claim \ref{thm:cl2}.}

Two remarks are in order. Firstly, our results hold also for very weak or even
zero coupling, i.e. when $\alpha\rightarrow0$, see the discussion in the last
paragraph of Section \ref{sec:final}. The same conclusion also holds for
exponential potentials, see Section \ref{sec:exp}. Secondly, the asymptotic
state when $\Omega_{k}\rightarrow1\ $and $\Omega_{\phi},\,\Omega_{\rho
}\rightarrow0$ corresponds to a Milne universe, see also \cite{nmc}. In that
case the asymptotic value of the effective equation of state parameter is
$w=-1/3$.

\begin{figure}[th]
\centering
%center subfigures in float
\begin{subfigure}{0.49\textwidth}
\includegraphics[width=\hsize]{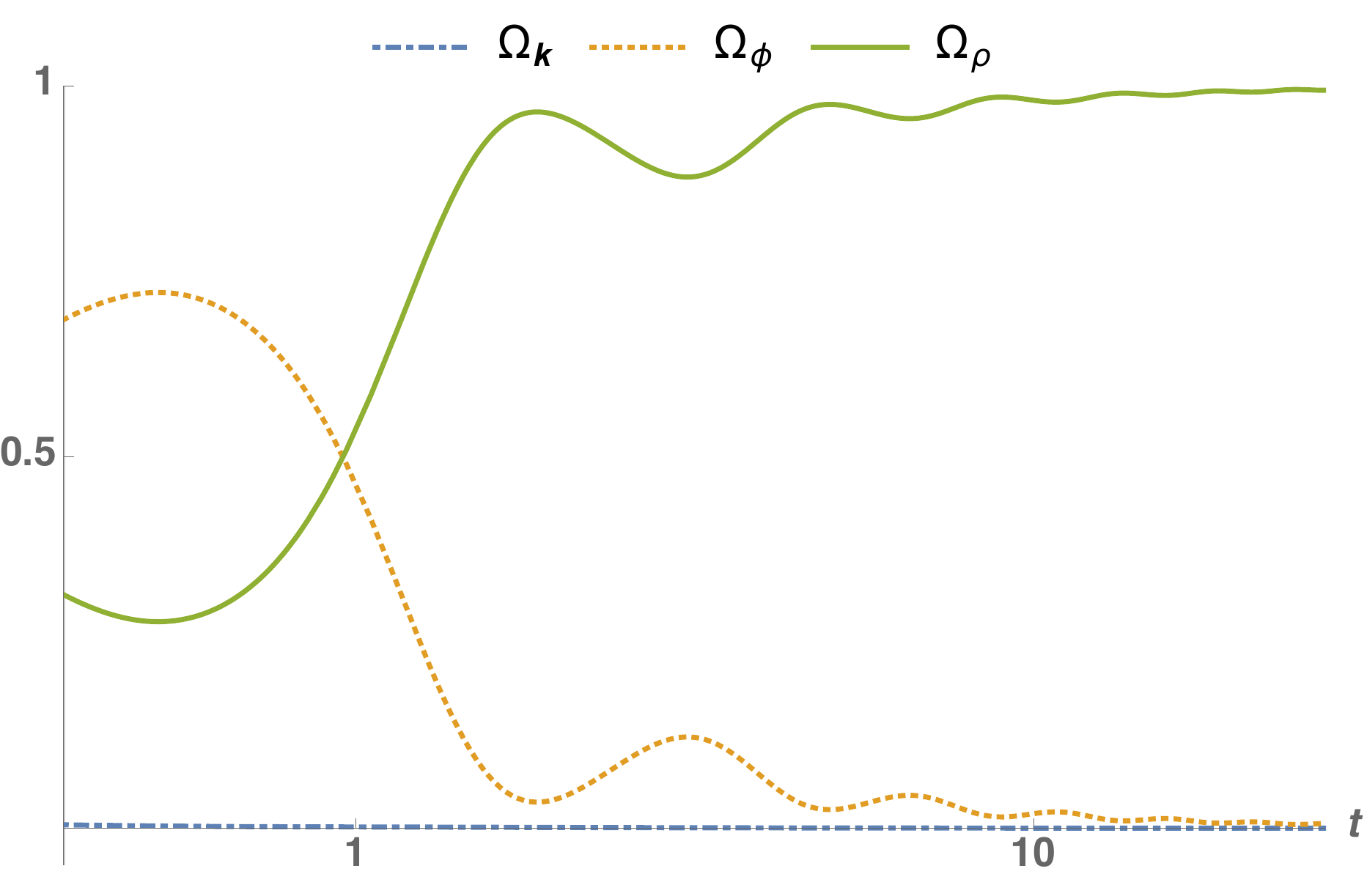}
\caption{$\gamma=1/3$}
\label{fig:en2}
\end{subfigure}
\hfil
%accomodate space between sub figures
\begin{subfigure}{0.49\textwidth}
\includegraphics[width=\hsize]{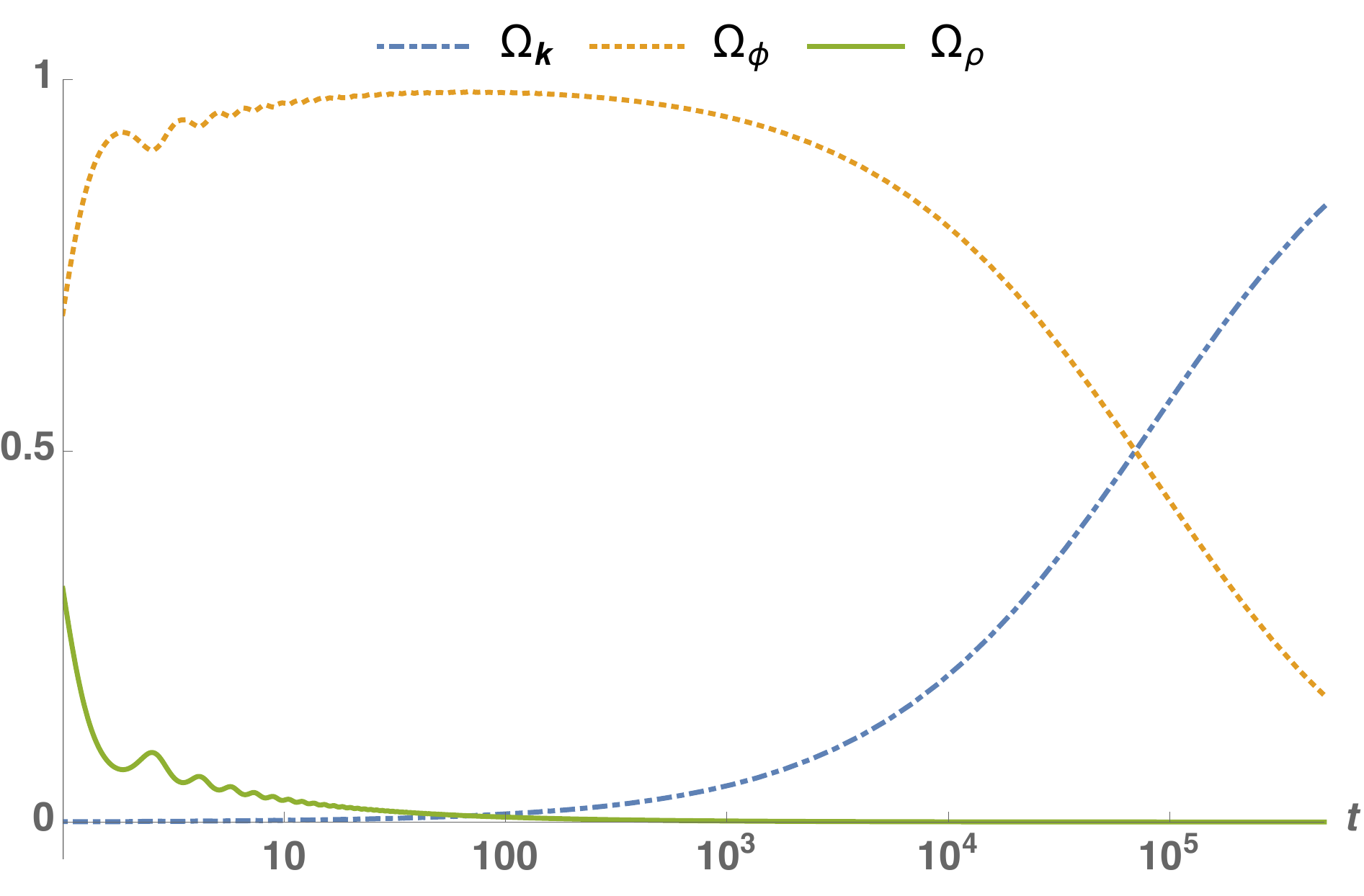}
\caption{$\gamma=4/3$}
\label{fig:en1}
\end{subfigure}
\caption{Normalized energy evolution with coupling function
$\chi\left(  \phi\right)  =e^{\sqrt{2/3}\phi}$, and potential $V(\phi
)=2\phi^{2}+o(\phi^{2})$, corresponding to the two different behaviors found
in Theorem \ref{final}.
 Initial data for the normalized energies are $\Omega_{\rho}=0.3145$, $\Omega_{\phi}=0.685$ and $\Omega
_{k}=0.0005$, within the error range expected in \cite{plco}.}
\label{fig:1}.
\end{figure}

\begin{figure}[th]
\centering
%center subfigures in float
\begin{subfigure}{0.49\textwidth}
\includegraphics[width=\hsize]{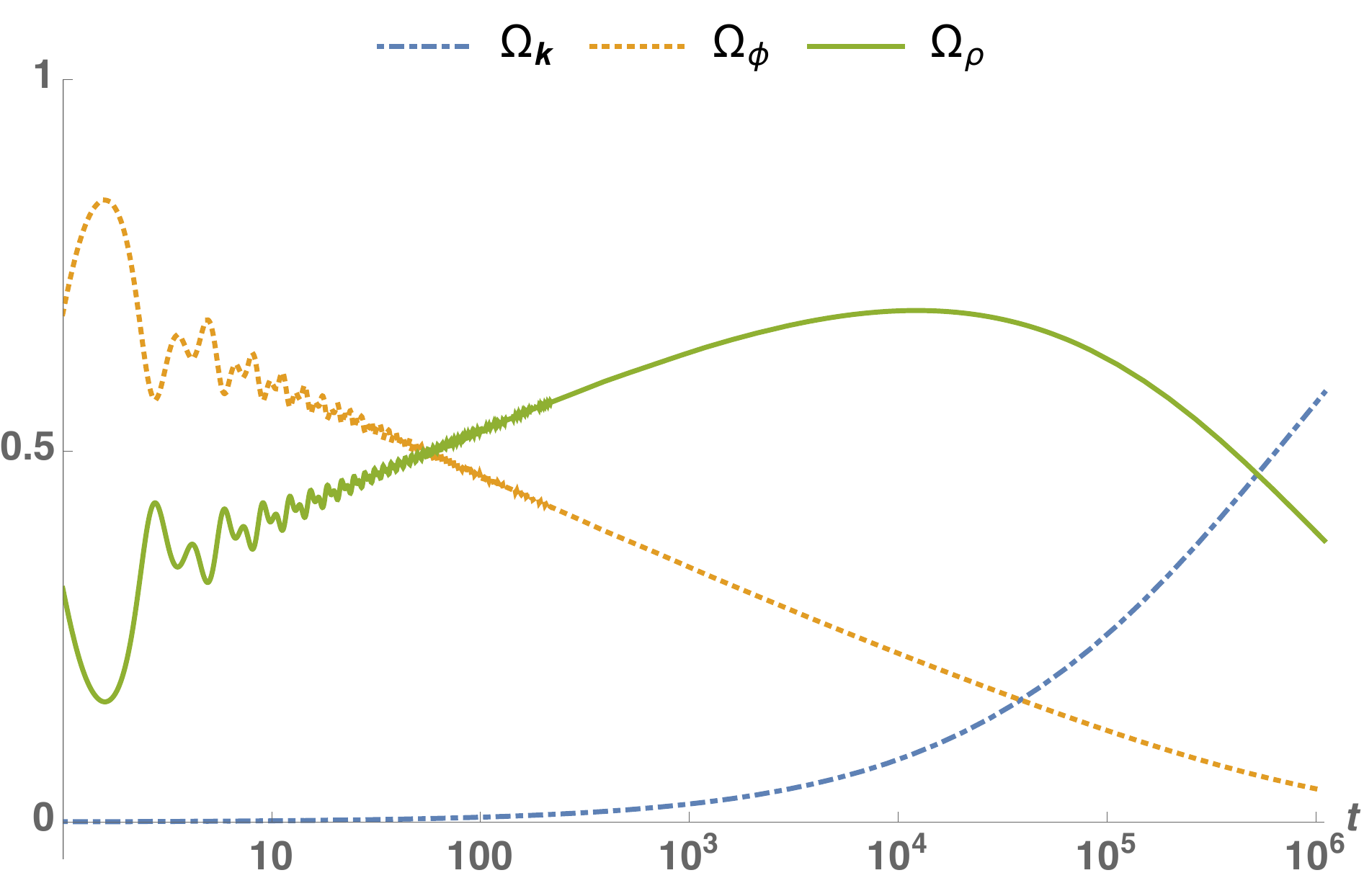}
\caption{$\gamma=0.9$}
\label{fig:en09}
\end{subfigure}
\hfil
%accomodate space between sub figures
\begin{subfigure}{0.49\textwidth}
\includegraphics[width=\hsize]{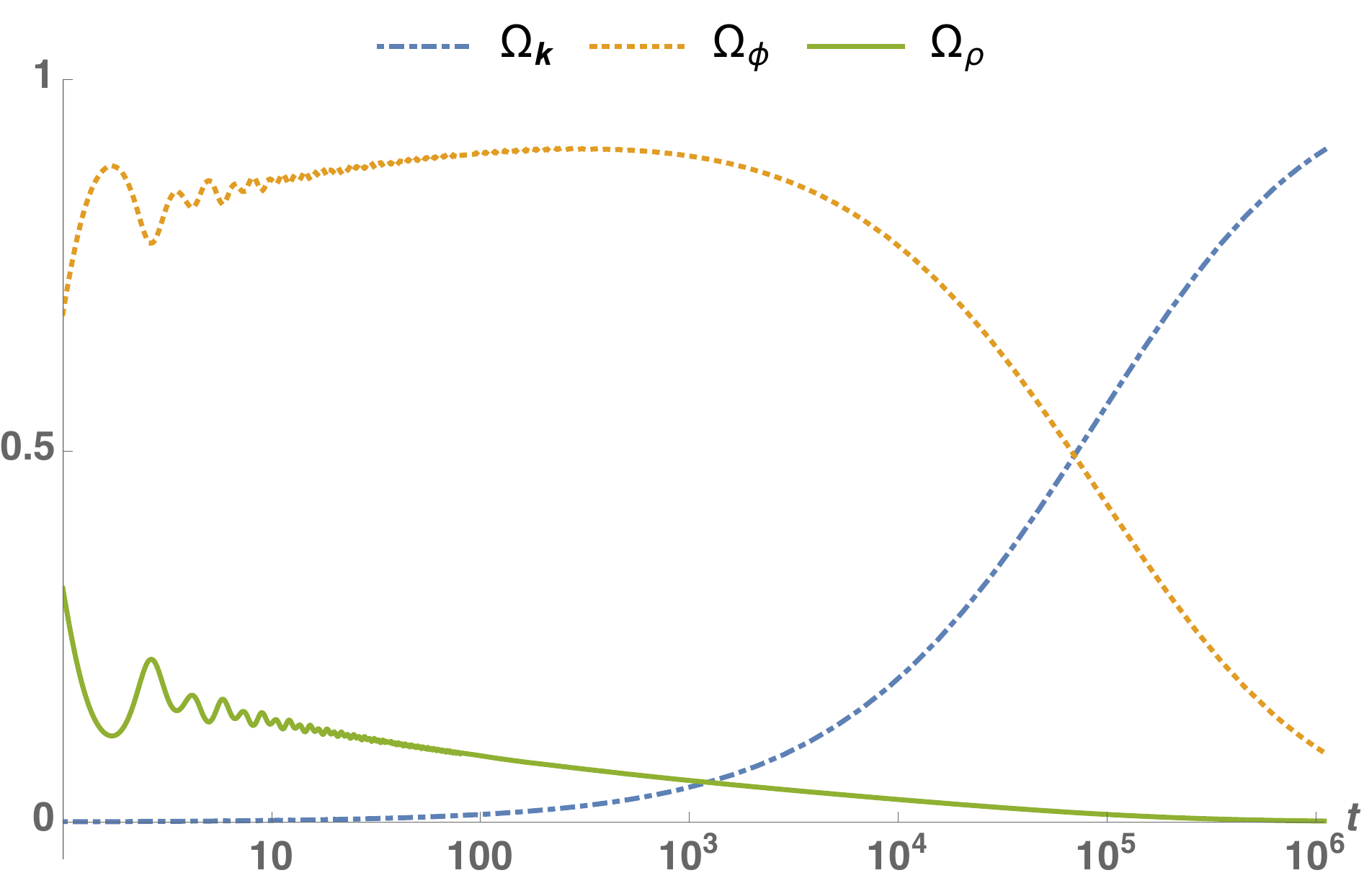}
\caption{$\gamma=1.1$}
\label{fig:en11}
\end{subfigure}
\caption{Normalized energy evolution for two \textquotedblleft
near-dust\textquotedblright cases. Here the  initial data, coupling
function and potential are the same as in Figure \ref{fig:1}, and only the value of $\gamma$ changes.}%
\label{fig:endust}%
\end{figure}

The different evolutions described in {Theorem \ref{final}} are illustrated in
Figures \ref{fig:1} and \ref{fig:endust}. In all cases we have considered a
coupling function $\chi\left(  \phi\right)  =e^{\sqrt{2/3}\phi}$, that gives
$\alpha$ constant. The constant value for $\lambda$ in potential
\eqref{eq:quadr-pot} has been picked up equal to 2. Moreover initial data have
been chosen in such a way that at some initial time, the universe expansion is
accelerated in accordance with \cite{plco}. 
Figure \ref{fig:en2} shows the
evolution of the energies in case $\gamma=1/3$: curvature energy remains small
and decreasing to zero, whereas matter energy rapidly dominates over
$\Omega_{\phi}$.
%In this case the acceleration of the
%universe remains positive throughout the evolution

The situation is completely different when $\gamma>2/3$. In Figure
\ref{fig:en1} is represented the case $\gamma=4/3$: in this case the curvature
energy slowly increases and eventually takes over the other energies, with
$\Omega_{\rho}$ rapidly vanishing and $\Omega_{\phi}$ decreasing in a more
persistent way.
%Also notice that, even starting with
%positive acceleration, the expansion changes to a phase of negative
%acceleration, and so on, eventually oscillating from one phase to another, see
%Figure \ref{fig:acc1}.

Of course, the way both scalar and matter energies decrease depends on
$\gamma$: see for instance Figure \ref{fig:endust} that represents the energy
distributions from the same initial data for two cases where pressures are
close to zero. As one would expect, $\Omega_{\rho}$ and $\Omega_{\phi}$ go to
zero keeping the mutual hierarchy of the flat case $k=0$ which depends on the
sign of $\gamma-1$, see \cite[Theorem 2]{gimi}.

\subsection{Exponential potential}

\label{sec:exp}

To broaden the analysis, let us consider a potential of the form
$V(\phi)=V_{0}e^{-\lambda\phi}+\ell$ and assume that there exists a constant
$\alpha_{0}$ such that
\begin{equation}
\alpha_{0}=\lim_{\phi\rightarrow+\infty}\alpha(\phi)=\frac{4-3\gamma}{2}%
\lim_{\phi\rightarrow+\infty}Q(\phi). \label{eq:assumQ}%
\end{equation}
As remarked in \cite{gimi} after the statement of Proposition 1, there is a
critical point \textquotedblleft at infinity\textquotedblright\ which is
asymptotically stable. If the critical value $\ell$ is strictly positive then
the scalar normalized energy approaches 1, against $\Omega_{\rho}$ and
$\Omega_{k}$ which go to zero. The subtle case is then to investigate what
happens when $\ell=0$, when $\phi\rightarrow+\infty$, and $y,\rho\rightarrow
0$. Expansion-normalized variable change
\begin{equation}
u=\frac{\sqrt{W}}{H},\quad w=\sqrt{\frac{V_{0}}{3}}\frac{e^{-\frac{\lambda}%
{2}\phi}}{H},\quad z=\frac{y}{\sqrt{6}H}, \label{eq:normvar}%
\end{equation}
together with the new time variable defined by $\mathrm{d}\tau=3H\mathrm{d}t$
bring system \eqref{Sistema} to the form%
\begin{equation}
\begin{aligned} u^{\prime} & =-\frac{1}{6}u\left( -3\gamma+(3\gamma-2)u^{2}+3\gamma w^{2}+3\gamma z^{2}-6z^{2}+2\right) ,\\ w^{\prime} & =-\frac{1}{6}w\left( -3\gamma+(3\gamma-2)u^{2}+3\gamma w^{2}+3\gamma z^{2}+\sqrt{6}\lambda z-6z^{2}\right) ,\\ z^{\prime} & =z\left( -1+z^{2}+\frac{u^{2}}{3}+\frac{\gamma}{2}\left( 1-u^{2}-w^{2}-z^{2}\right) \right) +\frac{1}{\sqrt{6}}\left( \lambda w^{2}+\alpha\left( 1-u^{2}-w^{2}-z^{2}\right) \right) ,\\ H^{\prime} & =-H\left( \frac13(u^2+3z^2)+\frac\gamma 2\left(1-u^2-w^2-z^2\right) \right). \end{aligned} \label{eq:systemex-exp}%
\end{equation}

Here a prime ($^{\prime}$) denotes differentiation with respect to the new
time $\tau$. In this case the evolution equation for $H$ decouples from the
rest of the evolution equations and so we have a system of three equations in
the unknowns $u,\,w,\,z$. The definition (\ref{eq:normvar}) implies that the
variables $u,\,w,\,z$ are all positive, included $z$; indeed we know that
$\phi\rightarrow+\infty$, therefore situations where $y$ is eventually
negative are irrelevant for our study.

A careful inspection of the asymptotically stable equilibria $(u_{\infty
},w_{\infty},z_{\infty})$ of \eqref{eq:systemex-exp} gives the following five
mutual exclusive situations, depending on the values of the three parameters
$\lambda>0,\gamma\in(0,2),\alpha_{0}\in\mathbb{R}$:
\begin{multline}
\mathcal{P}_{1}=\big(0,\sqrt{1-\frac{\lambda^{2}}{6}},\frac{\lambda}{\sqrt{6}%
}\big),\,\mathcal{P}_{2}=\big(\frac{\sqrt{\lambda^{2}-2}}{\lambda},\frac
{2}{\sqrt{3}\lambda},\frac{\sqrt{\frac{2}{3}}}{\lambda}\big),\,\mathcal{P}%
_{3}=\big(0,0,\frac{\sqrt{\frac{2}{3}}\alpha_{0}}{2-\gamma}%
\big),\label{eq:equil}\\
\mathcal{P}_{4}=\big(\frac{\sqrt{\alpha_{0}^{2}-\frac{3\gamma^{2}}{2}%
+4\gamma-2}}{\alpha_{0}},0,\frac{2-3\gamma}{\sqrt{6}\alpha_{0}}%
\big),\,\mathcal{P}_{5}=\big(0,\frac{\sqrt{\alpha_{0}^{2}-\alpha_{0}%
\lambda-\frac{3}{2}(\gamma-2)\gamma}}{\lambda-\alpha_{0}},\frac{\sqrt{\frac
{3}{2}}\gamma}{\lambda-\alpha_{0}}\big).
\end{multline}
Recalling \eqref{eq:energyuwz}
it is easy to see that in cases $\mathcal{P}_{3}$ and $\mathcal{P}_{5}$ either
the scalar energy or the fluid energy eventually dominate, in case
$\mathcal{P}_{1}$ the scalar energy eventually totally dominates, whereas in
cases $\mathcal{P}_{2}$ and $\mathcal{P}_{4}$ there might be situations where
the curvature energy $\Omega_{k}$ eventually dominates.

\begin{figure}[th]
\centering
%center subfigures in float
\includegraphics[width=0.5\hsize]{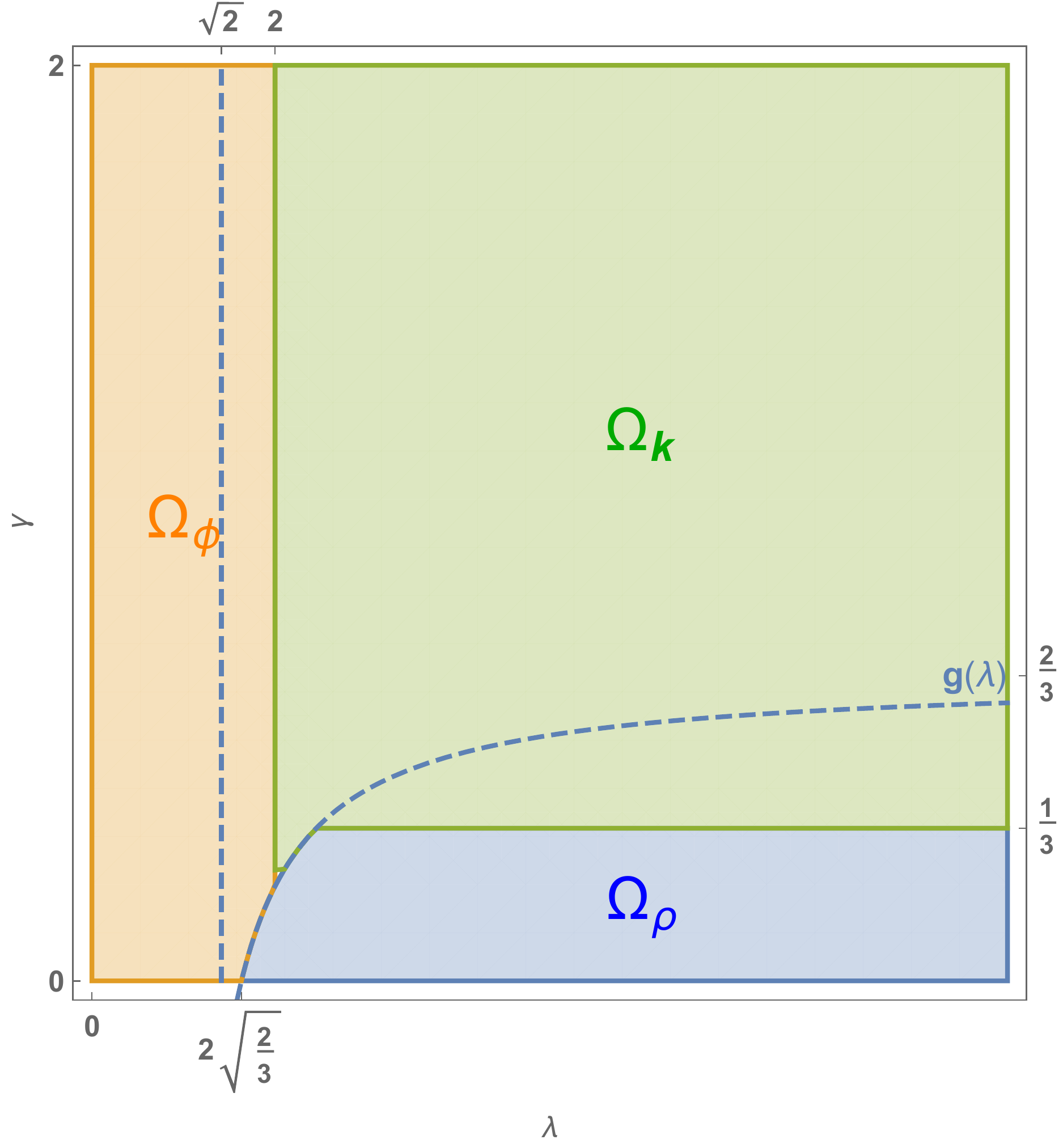} \caption{Asymptotically
dominating energy when $V=V_{0}e^{-\lambda\phi}$ and the coupling function is
given by $\chi(\phi)=e^{\sqrt{2/3}\phi}$, in terms of the two free parameters
$\lambda$, $\gamma$. The dashed lines separate the regions where the system
\eqref{eq:systemex-exp} approaches the three possibile cases $\mathcal{P}_{1}%
$, $\mathcal{P}_{2}$ and $\mathcal{P}_{4}$ ($H\rightarrow0^{+}$ always).}%
\label{fig:exp}%
\end{figure}

Moreover, the scale factor behaves for $t\rightarrow+\infty$ as
\[
a(t)\approx t^{p},\qquad p=\left(  u_{\infty}^{2}+3z_{\infty}^{2}+\frac{3}%
{2}\gamma\left(  1-u_{\infty}^{2}-w_{\infty}^{2}-z_{\infty}^{2}\right)
\right)  ^{-1}.
\]

As an example, let us give the detailed results when $\chi(\phi)=e^{\sqrt
{2/3}\phi}$, so that $\alpha=\alpha_{0}=\left(  4-3\gamma\right)  \sqrt{6}$.
In this case only the three cases $\mathcal{P}_{1}$, $\mathcal{P}_{2}$ and
$\mathcal{P}_{4}$ may take place, respectively:

\begin{itemize}
\item $\mathcal{P}_{1}$: $0<\lambda<\sqrt{2}$, and $a(t)\approx t^{2/\lambda
^{2}}$;

\item $\mathcal{P}_{2}$: $\lambda>\sqrt{2}$ and $\mathrm{max}\left\{
0,g(\lambda)\right\}  <\gamma<2$, where $g(\lambda)=\frac{6\lambda-4\sqrt{6}%
}{9\lambda-3\sqrt{6}}$, and $a(t)\approx t$;

\item $\mathcal{P}_{4}$: $\lambda>2 \sqrt{\frac{2}{3}}$, $0<\gamma<g(\lambda
)$, and $a(t)\approx t$.
\end{itemize}

Using the fourth equation in \eqref{eq:systemex-exp} we find
\[
\ddot{a}=a\left(  H^{2}+3HH^{\prime}\right)  =1-u^{2}-3z^{2}+\frac{3}{2}%
\gamma\left(  u^{2}+w^{2}+z^{2}-1\right)  ,
\]
and therefore examples with positive acceleration at initial data can be
easily provided.

Figure \ref{fig:exp} represent the strip $(\lambda,\gamma)\in(0,+\infty
)\times(0,2)$: the dashed lines separate the three regions described above.
Using \eqref{eq:energyuwz} we conclude that there is a wide region of the two
parameters where the curvature energy dominates over the other two energies. 

\section{Further discussion}

\label{sec:final}

It is well known that astronomical measurements constrain the spatial
curvature to be very close to zero, although they do not constrain its sign
\cite{K19}. For example, results of the Planck mission released in 2015 show
the cosmological curvature parameter, $\Omega_{k}$, to be $0.000\pm0.005$,
consistent with a flat universe. More recent constraints on $\Omega_{k}$,
yield the values $0.001\pm0.002$, \cite{plco}. Similar bounds for $\Omega_{k}$
were found in an analysis of 42 measurements of the Hubble parameter and
baryon acoustic oscillation data in \cite{rdr}, although the authors stress
that \textquotedblleft more and better data are needed before we can make
definitive statements about the spatial curvature of the
universe\textquotedblright. Larger bounds $0.09\pm0.19$ for $\Omega_{k}$ were
found in a model-independent method to test the curvature of the universe
\cite{yuwa}. Although a flat universe is usually assumed when one studies the
nature of dark energy \cite{Ry}, for some models of dark energy, an open
universe is more favoured than a flat universe \cite{ikst}. Thus a flat
universe may not be a good assumption for constraining some particular models
of dark energy.

In this investigation we analysed the late time evolution of negatively curved
expanding FLRW models having a scalar field coupled to matter. For exponential
potentials $V\left(  \phi\right)  =V_{0}e^{-\lambda\phi}$, the energy density
of the scalar curvature eventually dominates over both the perfect fluid and
the scalar field for a wide range of the parameters $\lambda$ and $\gamma$.
For non--negative local minima of $V$, the corresponding equilibria share the
same properties as in the flat case \cite{gimi} namely, they are
asymptotically stable and in case the minimum is positive, say $V(\phi_{\ast
})>0$, the energy density of the scalar field eventually rules over the energy
density of the fluid and the asymptotic state has an effective cosmological
constant $V(\phi_{\ast})$. In that case, the equilibrium solution represents
an accelerating future attractor. In case the minimum is zero and
nondegenerate, then $\Omega_{\rho}$ eventually dominates if $\gamma<2/3$.
However, if $\gamma>2/3$, the energy density of the scalar curvature
eventually dominates over both the perfect fluid and the scalar field.

One can understand this result by the following heuristic reasoning.
Integrating the energy density equation (\ref{conssfjm}), we obtain $\rho=$
$ce^{-\alpha\phi}a^{-3\gamma}$. Since $\phi\rightarrow\phi_{\ast}$ as
$t\rightarrow\infty$, absorbing $e^{-\alpha\phi_{\ast}}$ into the constant of
integration, we can write $\rho\simeq ca^{-3\gamma}$ as $t\rightarrow\infty$.
If one could use the Kryloff-Bogoliuboff (KB) approximation \cite{krbo} it
could be shown that near the equilibrium the energy density of the scalar
field decreases as $\epsilon\simeq a^{-3}$. Since $H^{2} \Omega_{k}$ scales as
$a^{-2},$ it is reasonable to conclude that for $\gamma>2/3$ it dominates over
both $\epsilon$ and $\rho$ at late times. As remarked in \cite{migi}, the main
obstruction in applying KB approximation lies in the impossibility to have an
a--priori estimate on $\rho/\epsilon=\Omega_{\rho}/\Omega_{\phi}$, therefore
the need of alternative strategies to attack the problem.

The above results were rigorously proved assuming only that critical points
are finite and at those points $V(\phi)$ is non-negative. No further
assumption on $V$ enter in the study of the late time behavior around a
critical point $\phi_{\ast}\in\mathbb{R}$, because for that situation only the
behavior of the potential near $\phi_{\ast}$ is important and no growth at
infinity assumptions on $V$ are actually needed.

A similar behavior as in Theorem \ref{final} of the curvature over two
interacting fluids was observed in \cite{nmc} for $k=-1$ models. Recalling
that in the massless case the scalar field reduces to a stiff fluid, in view
of our result we can say that the $\gamma=2/3$ threshold found in \cite{nmc}
is stable with respect to the interaction with a potential having a local
minimum with vanishing critical value. The curvature dominated asymptotic
state was also found in \cite{hcw} where FLRW models with an exponential
potential and a barotropic fluid without coupling were studied. Nonflat
cosmologies have been analytically studied also in \cite{K6}, where dark
energy is modeled by a cosmological constant and the problem is recast into a
competitive species dynamical framework.

An important question that should be further investigated is the case of
closed cosmologies \cite{ra2}. Based on the experimentation with the potential
$V\left(  \phi\right)  =V_{0}\left(  1-e^{-\sqrt{2/3}\phi}\right)  ^{2}$ which
arises in the conformal frame of quadratic gravity \cite{miri}, we believe
that a closed model cannot avoid recollapse, unless the minimum of the
potential is strictly positive (see \cite{cogo} for non interacting fluid and
scalar field). In that case, the asymptotic state must be de Sitter space.

An other important issue is the strength of the coupling function $Q\left(
\phi\right)  $. For viable dark energy models, it is necessary that the energy
density of the scalar field remains insignificant during most of the history
of the universe and emerges only at late times to account for the current
acceleration of the universe. However, in models with double exponential
potentials it was observed that only a very weak coupling of the scalar field
to ordinary matter can lead to acceptable cosmological histories of the
universe \cite{tzmi}. This fact reinforces the general conclusions in
\cite{apt,agpt,amen2}, that HOG dark energy models with $f(R)=R-\mu
^{2(n+1)}/R^{n}$, where $\mu>0,n>1$, are not cosmologically viable. This
result is attributed to the fact that in these theories, matter is strongly
coupled to gravity (recall that $Q=\sqrt{2/3}$ in HOG theories). Nevertheless,
in \cite{cmr,cnot} specific examples of $f(R)\sim R^{n}$ gravity models were
built, including matter and accelerated phases which are cosmological viable,
at the expense of having noninteger values for $n$. We do not enter into the
old discussion about the equivalence issue of the Einstein and the Jordan
frame (see for example \cite{bran}; see also \cite{cmr1} with specific
examples and the extended review articles \cite{sofa,fets,od1,cala,od2} with
references therein). As mentioned after the proof of Theorem \ref{final}, our
results hold for every value of the coupling function, even for $\alpha
\rightarrow0$. Had we allowed a dynamical role to $Q$, it would be very
interesting to see if the dynamics leads to a very tiny value of $Q$ at late
times. Such a result could lead to a generalization of the attractor mechanism
of scalar-tensor theories towards general relativity, found by Damour and
Nordtvedt in the case of a massless scalar field \cite{dano,minu}.

\section*{Acknowledgement}

The authors wish to thank Salvatore Capozziello and Orlando Luongo for
valuable discussions and suggestions. We also thank an anonymous referee for
his suggestions which helped us to clarify some points.


\begin{thebibliography}{99}                                                                                               %
\rhead[\fancyplain{}{\bfseries \leftmark}]{\fancyplain{}{\bfseries
\thepage}}

\bibitem{K3} A.A. Coley, \textit{Dynamical Systems and Cosmology}, (Kluwer
Academic Publishers, Dordrecht, Boston, London, 2003).

\bibitem{wein2} S. Weinberg, \textit{Cosmology}, (OUP, Oxford, 2008).

\bibitem{K11} S. Bahamonde, C.G. B\"{o}hmer, S. Carloni, E.J. Copeland, W.
Fang, N. Tamanini, Phys. Rep. \textbf{775--777}, 1--122, (2018).

\bibitem{gasp} M. Gasperini, \textit{Elements of String Cosmology}, (CUP,
Cambridge, 2007).

\bibitem{fuma} Y. Fuji, K. Maeda, \textit{The Scalar-Tensor Theory of
Gravitation}, (CUP, Cambridge, 2003); V. Faraoni, \textit{Cosmology in
Scalar-Tensor Gravity},  (Kluwer Academic Publishers, Dordrecht, 2004).

\bibitem{K14} R. Bean, D. Bernat, L. Pogosian, A. Silvestri, M. Trodden,
Phys. Rev. D \textbf{75} (6), 064020, (2007).

\bibitem{lsf} G. Leon, P. Silveira, C.R. Fadragas, arXiv:1009.0689 [gr-qc]
(2010).

\bibitem{khwe} J. Khoury, A. Weltman, Phys. Rev. D \textbf{69}, 044026,
(2004).

\bibitem{wate} T.P. Waterhouse, arXiv:astro-ph/0611816 (2006).

\bibitem{amen1} L. Amendola, Phys. Rev. D \textbf{62}, 043511, (2000).

\bibitem{psc} A. Pourtsidou, C. Skordis, E.J. Copeland, Phys. Rev. D \textbf{%
88,} 083505, (2013); E.J. Copeland, A.R. Liddle, D. Wands, Phys. Rev. D 
\textbf{57}, 4686, (1988).

\bibitem{lumu} O. Luongo, M. Muccino, Phys Rev D \textbf{98}, 103520, (2018).

\bibitem{bico} A.P. Billyard, A.A. Coley, Phys. Rev. D \textbf{61}, 083503,
(2000).

\bibitem{fara1} V. Faraoni, Phys. Rev. D \textbf{62}, 023504, (2000).

\bibitem{wael} J. Wainwright, G.F.R. Ellis, \textit{Dynamical Systems in
Cosmology}, (CUP, Cambridge, 1997).

\bibitem{bkgz} V.A. Belinsky, I.M. Khalatnikov, L.P. Grishchuk, Y.B.
Zeldovich, Phys. Lett. B \textbf{155}, 232, (1985).

\bibitem{K9} Wei Fang, Ying Li, Kai Zhang, Hui-Qing Lu, Class. Quantum Grav. 
\textbf{26}, 155005, (2009).

\bibitem{gimi} R. Giamb\`{o}, J. Miritzis, Class. Quantum Grav. \textbf{27},
095003, (2010).

\bibitem{dkzpct} N. Dimakis, A. Karagiorgos, A. Zampeli, A. Paliathanasis,
T. Christodoulakis, P.A. Terzis, Phys. Rev. D \textbf{93}, 123518, (2016).

\bibitem{fost} S. Foster, Class. Quant. Grav. \textbf{15}, 3485, (1998).

\bibitem{miri03} J. Miritzis, Class. Quantum Grav. \textbf{20}, 2981, (2003).

\bibitem{nmc} A. Nunes, J.P. Mimoso, T.C. Charters, Phys. Rev. D \textbf{63}%
, 083506, (2001).

\bibitem{plco} N. Aghanim et al, Planck Collaboration, (arXiv:1807.06209
[astro-ph.CO]) (2018).

\bibitem{K19} M. Kowalski et al, The Supernova Cosmology Project, ApJ 
\textbf{686}, 749--778, (2008).

\bibitem{rdr} J. Ryan, S. Doshi, B. Ratra, Mon. Not. Roy. Astron. Soc. 
\textbf{480} no.1, 759--767, (2018).

\bibitem{yuwa} H. Yu, F.Y. Wang, ApJ \textbf{828}, 85, (2016).

\bibitem{Ry} G. Ryskin, Astropart. Phys. \textbf{62}, 258--268, (2015).

\bibitem{ikst} K. Ichikawa, M. Kawasaki, T. Sekiguchi, T. Takahashi, J.
Cosmol. Astropart. Phys. \textbf{12}, 005, (2006).

\bibitem{krbo} N. Kryloff, N. Bogoliuboff, \textit{Introduction to Nonlinear
Mechanics}, (PUP, Princeton, 1943).

\bibitem{migi} J. Miritzis, R. Giamb\`{o}, \textit{AIP Conf. Proc}. \textbf{%
1241}, 1061, (2010).

\bibitem{hcw} R.J. van den Hoogen, A.A. Coley, D. Wand, Class. Quantum Grav. 
\textbf{16},1843, (1999).

\bibitem{K6} J. Perez, A. F\"{u}zfa, T. Carletti, L. M\'{e}lot, L.
Guedezounme, Gen. Rel. Grav. \textbf{46}, 1753, (2014).

\bibitem{ra2} C.G. Park, B. Ratra, ApJ \textbf{868}, 83, (2018).

\bibitem{miri} J. Miritzis, J. Math. Phys. \textbf{46}, 082502, (2005).

\bibitem{cogo} A.A. Coley, M. Goliath, Phys. Rev. D \textbf{62}, 043526,
(2000).

\bibitem{tzmi} K. Tzanni, J. Miritzis, Phys. Rev. D \textbf{89}, 103540,
(2014).

\bibitem{apt} L. Amendola, D. Polarski, S. Tsujikawa, Phys. Rev. Lett.%
\textbf{\ 98}, 131302, (2007).

\bibitem{agpt} L. Amendola, R. Gannouji, D. Polarski, S. Tsujikawa, Phys.
Rev. D \textbf{75}, 083504, (2007).

\bibitem{amen2} L. Amendola, D. Polarski, S. Tsujikawa, Int. J. Mod. Phys. D 
\textbf{16}, 1555, (2007).

\bibitem{cmr} S. Capozziello, P. Martin-Moruno, C. Rubano, Phys. Lett. B 
\textbf{664}, 12--15, (2008).

\bibitem{cnot} S. Capozziello, S. Nojiri, S.D. Odintsov, A. Troisi, Phys.
Lett. B \textbf{639}, 135--143, (2006).

\bibitem{bran} M. Ferraris, M. Francaviglia, G. Magnano, Class. Quantum
Grav. \textbf{7}, 261, (1990); S. Cotsakis, Phys. Rev. D \textbf{47}, 1437,
(1993); Erratum Phys. Rev. D \textbf{49}, 1145, (1994); S. Capozziello, R.
de Ritis, A.A. Marino, Class. Quantum Grav. \textbf{14}, 3243, (1997); C.H.
Brans, Class. Quantum Grav. 5 L, 197, (1998); G. Magnano, L.M. Soko\l owski,
Phys. Rev. D \textbf{50}, 5039, (1994); V. Faraoni, E. Gunzig, P. Nardone,
Fund. Cosmic Phys. \textbf{20}, 121, (1999); V. Faraoni, S. Nadeau, Phys.
Rev. D \textbf{75}, 023501, (2007); S. Capozziello, M. Francaviglia, Gen.
Relativ. Gravit. \textbf{40}, 357--420,  (2008); G. Leon, Class. Quantum
Grav. \textbf{26}, 035008, (2009).

\bibitem{cmr1} S. Capozziello, P. Martin-Moruno, C. Rubano, Phys. Lett. B 
\textbf{689}, 117--121, (2010).

\bibitem{sofa} T. Sotiriou, V. Faraoni, Rev. Mod. Phys. \textbf{82},
451-497, (2010).

\bibitem{fets} A. De Felice, S. Tsujikawa, Living Rev. Relativity \textbf{13}%
, 3-161, (2010).

\bibitem{od1} S. Nojiri, S.D. Odintsov, Phys. Rep. \textbf{505}, 59--144,
(2011).

\bibitem{cala} S. Capozziello, M. De Laurentis, Phys. Rep. \textbf{509},
167--321, (2011).

\bibitem{od2} S. Nojiri, S.D. Odintsov, V.K. Oikonomou, Phys. Rep. \textbf{%
692}, 1--104, (2017).

\bibitem{dano} T. Damour, K. Nordtvedt, Phys. Rev. Lett. \textbf{70}, 2217,
(1993); T. Damour, K. Nordtvedt, Phys. Rev. D \textbf{48}, 3436, (1993).

\bibitem{minu} I.P. Mimoso, A. Nunes, Astrophys. Space Sci. \textbf{283} no.
4, 661, (2003).

\end{thebibliography}
\end{document}